# Effect of shaping plate apparatus on mechanical properties of 3D printed cement-based materials: Experimental and numerical studies

Tinghong Pan[a1], Huaijin Teng[b1], Hengcheng Liao[b*], Yaqing Jiang[a*], Chunxiang Qian[b], Yu Wang[a]

[a] *College of Mechanics and Materials, Hohai University, Nanjing 211100, China;*

[b]*School of Materials Science and Engineering, Southeast University, Nanjing 211189, China*

[*]*corresponding author*

*E-mail addresses: hengchengliao@seu.edu.cn (H. Liao), yqjiang@hhu.edu.cn (Y. Jiang).*

[1]*co-first authors: These authors contributed to the work equally and should be regarded as co-first authors.*

**Abstract:**

Precisely controlling the shape of the printed-layers, eliminating the curved sides and internal stress concentration, and increasing the mechanical properties are essential to guarantee the quality of 3D printed cement-based structures. This work aims at achieving the above-mentioned targets through a specially designed shaping plate apparatus. The pressure (stress) distribution in the printed structure with a shaping plate apparatus (SP-3DPC), and the cross-sectional shape, microstructure and mechanical properties of SP-3DPC were systematically investigated. Results indicate that using the shaping plate apparatus may slightly reduce the printing speed, but it can effectively constrain the free expansion of extrudate, control its cross-sectional geometry, and improve the surface finish quality and mechanical properties of the printed structure. This study provides a theoretical basis and technical guidance for the design and application of the shaping plate apparatus.

---

[*] Corresponding authors.
E-mail addresses: hengchengliao@seu.edu.cn (H. Liao), yqjiang@hhu.edu.cn (Y. Jiang),.

## 1. Introduction

3D printing technology for cementitious material (3DPC) newly emerges in construction industry [1-3], which combines additive manufacturing (AM) process with cement-based materials. It has a potential to improve the automation level of building structure manufacturing [4-6]. In the past few years, 3DPC technology has been developed rapidly, which has brought many significant technological advancements for building-construction industry, such as not needing the formworks [7, 8], reducing the design time, execution the period and labor costs [9-11], reducing the wastes and increasing the design-freedom [8, 10]. 3DPC technology involved in this work mainly refers to the layered extrusion 3DPC one, which have relied on the sequential construction of a geometry by stacking one layer over the previous one. Selecting binder jetting 3DPC technology [12-14] will not be covered in this work.

In the extrusion-deposition process of cement-based materials, the printing ink was extruded from one or a few nozzles at a constant flow rate, and then deposited layer−by−layer through a digitally controlled robot [15, 16]. Ignoring the case that the flow behaviour of extrudate in deposition process is constrained in some directions via trowels, two asymptotic regimes ("infinite brick extrusion" regime and "free flow deposition" regime) are proposed to describe the shape formation after the deposition process [17, 18], as shown in Figure 1. In the first case (called "infinite brick extrusion" regime), the printing ink possess higher yield stress than the stress induced by gravity and/or resulting from pumping pressure, and the ink is unsheared before extrusion. The material out of the nozzle as a stiff continuous filament, and the cross-sectional shape of the printed filament is mainly imposed by the geometry of nozzle opening [16, 19, 20]. However, the higher yield stress or stiffness of the printing ink may bring challenges to the extrusion process and the fusion behavior between adjacent layers [21, 22]. In the second case (called "free flow deposition" regime), the material is fully sheared before extrusion, due to the local screw mixer or local contraction [18, 19]. This situation is similar to traditional extrusion-deposition process. When the sheared material is deposited on the build surface or previous printed layer, it will result in serious deformation and surface quality problems. Therefore, curved side of layer and staircase effect are typically observed in 3DPC structures, due to the free expansion and deformation of the printed filament. This may significantly affect the mechanical properties of the printed structure, due to the severe stress concentration at interface [23-25]. In this regime, the final geometry of the printed filament is expected to result from the competition between gravity and yield stress [16]. The transition between these two regimes depends on the rheological properties of the printing ink, on the process parameters (such as flow rate, printing speed or nozzle stand-off distance) and on the nozzle geometry. How to precisely control the shape of layers, eliminate the curved side of layers and internal stress concentration and increase the mechanical properties are a very opened research field.

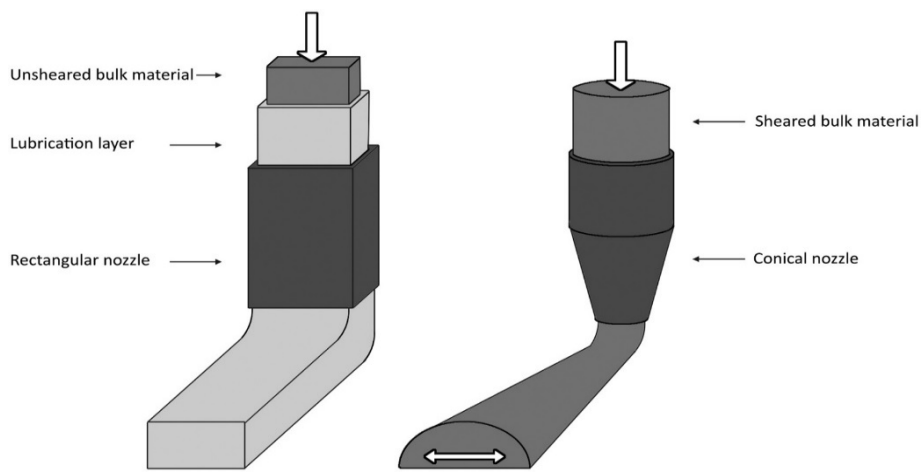

Figure 1. Two asymptotic regimes for concrete extrusion in additive manufacturing. Left: so-called "infinite brick" regime; Right: "free flow deposition" regime, extracted from [18].

The design of the printing head and nozzle is critical to better control the shape of the printed filament [18]. Generally, a round nozzle can provide the printing freedom and makes it easier to change the nozzle angle. However, the low percentage of contact area may affect the buildability and interlayer bonding strength of the printed structure [26, 27]. The square or rectangular nozzle can provide higher contact area between the adjacent printed layers, which can increase the interface stability and interlayer bonding strength of the printed structures [27]. However, Manikandan et al. [28] reported that the circular nozzle results in a higher surface finish quality, while the square nozzle imparts higher surface roughness and contour deviation in the printing of cylindrical constructs. The geometry of the above-mentioned nozzles is usually fixed, and thus the geometry of the extrudate cannot be adjusted to conform to the designed geometry of the printed structure [25]. However, the variable-geometry nozzle makes it possible to print structures with a smoother surface finish and a less obvious layer boundary, which will improve the surface quality and mechanical properties [24, 25]. The geometry of nozzles has an important influence on the surface quality and mechanical properties of the printed structures.

Furthermore, Khoshnevis and Carneau et al. [29-31] reported that trowels (i.e., shaping plate) can impose horizontal constraint on the extrudate during extrusion-deposition process, and hence enhance geometrical control of the printed filament. It can automatically adjust the cross section of the printed filament to fit the desired layer geometry [29, 30]. Automatically controlled multiple shaping plates ensure an accurate modelling of specific wall geometry [6], as shown in Figure 2. However, the influence of the shaping plate on the surface finish quality, mechanical properties and microstructure of 3D printed structures has not been well established. The mechanism of the shaping plate controlling the flow and deformation behaviors of extrudate is still unclear. Better understanding the mechanism of the shaping plate is of great significance to prepare more advanced printhead and develop 3DPC technology.

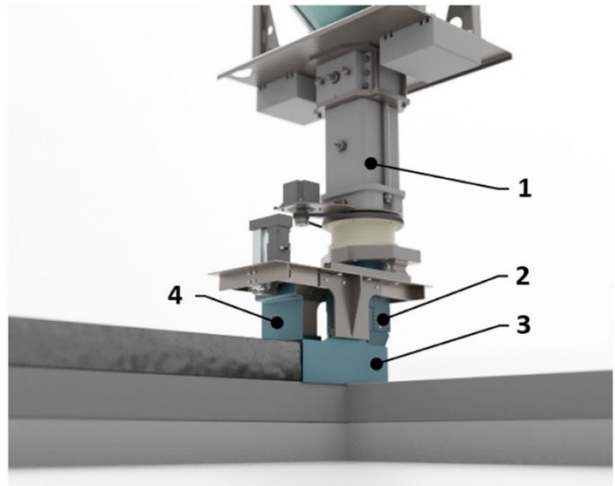

Figure 2. Printhead under development [6]. (1) extruder and the complex forming system with (2) nozzle, (3) shaping plates, (4) shutter and several actuators for printing various wall cross-sections.

In this paper, a shaping plate apparatus was specially designed to precisely control of the shape of the printed layers, eliminate the curved side and internal stress concentration, and increase the mechanical properties. The macro- and microscopic properties of the printed structures with and without the shaping plate apparatus were deeply investigated. The mechanical properties were measured through experiments. The microstructure was characterized via X-ray micro-computed tomography (CT). The stress (pressure) distributions were visualized and quantitatively characterized by single-phase computational fluid dynamics (CFD) simulation. This work systematically studies the pressure (stress) distribution in the printed structure with a shaping plate apparatus (SP-3DPC), cross-sectional shape, microstructure and mechanical properties of SP-3DPC, aiming to reveal the mechanism of the shaping plate apparatus and to provide a practical direction and a theoretical guidance for the further development of shaping plate apparatus and 3DPC technology.

## 2. Preparation of 3D printed cement-based materials

### 2.1. Materials

The adopted cementitious material for 3D printing in this study is Portland cement

(OPC, Type II, 52.5 grade, Nanjing Conch Cement Co. Ltd., Nanjing, China), and its chemical composition is listed in Table 1. The selected ground river-sand is with a fineness modulus of 2.3 and a maximum particle size of 1 mm, and the self-synthesized polycarboxylate superplasticizer (PCE) is with an effective solid content of 40 % and a water reducing rate of 30 %. Nano attapulgite (Jiangsu Jiuchuan Nano-material Technology Co.Ltd.) is used as thixotropic modifier, and hydroxy propyl methyl cellulose (HPMC, Renqiu Cheng Yi Chemical Co.Ltd.) as viscosity modifier.

Table 1. Chemical composition of the selected cement [wt.%].

| CaO | $SiO_2$ | $Al_2O_3$ | $Fe_2O_3$ | MgO | $SO_3$ | LOI | Total |
|---|---|---|---|---|---|---|---|
| 64.50 | 22.04 | 4.76 | 3.87 | 0.92 | 1.90 | 2.01 | 100 |

*2.2. Mixing ratio and mixing method*

In this work, the ratio of water to cement is set as 0.32; the ratio of aggregate to cementitious material is set as 1:1.5; the dosage of PCE is set as 3 ‰ (by mass) of the cementitious material, nano attapulgite is 8 ‰ (by mass) and HPMC is 3 ‰ (by mass).

The dry cement, fine sand, nano attapulgite and HPMC are weighed according to the proportion, and then are placed into a planetary cement mortar mixer (JJ-5) to mix evenly. Next, the water and PCE are weighed and poured into mixer at one time. The wet batch and dry batch are mechanically mixed for 60 s at a slow speed of 140 r/min and for 90 s at a high speed of 285 r/min, respectively, to form an evenly cement mortar. The prepared mixture is as the "mass source" of 3D printing.

*2.3. Rheological property characterization of 3D printed cement-based materials*

Rheological property is an important parameter affecting the printability of 3D printing cement-based materials. Many studies have illustrated that the fresh mortar (concrete) belongs to Bingham fluid, and Bingham rheological model can well describe the rheological behaviour of fresh concrete [9, 23]. The main rheological parameters are yield stress and plastic viscosity, which can be obtained from the up/down curves of the flow test in a rotational rheometer.

The controlled shear rate (CSR) protocol is used to obtain the viscosity and yield stress of the testing material. Before CSR test, the mixture should be pre-sheared for 120 s by applying a constant shear rate of 100 s$^{-1}$ to homogenize the sample. Then an increasing shear rate ramp from 0 s$^{-1}$ to 100 s$^{-1}$ within 100 s and a decreasing shear rate ramp from 100 s$^{-1}$ to 0 s$^{-1}$ with 100 s are successively applied to obtain the up-curve and down-curve of the flow test. The down-curve (as shown in Figure 3) is used to calculate the rheological parameters. The viscosity and yield stress are obtained by fitting the down-curve with the Bingham model, and the calculation range is in the stabilization stage of 20-80 s$^{-1}$.

As shown in Figure 3, the determination coefficient ($R^2$) between the down-curve and Bingham model is 0.96, indicating that using Bingham rheological model can well describe the constitutive behaviour of the cement mortar used in this work. The yield stress of the mortar corresponds to the vertical intercept (the shear stress value when shear rate is zero) of the fitted curve, while the viscosity corresponds to the slope. The viscosity and yield stress of the cement mortar used in this work are 4.83 Pa·s and 276 Pa, respectively.

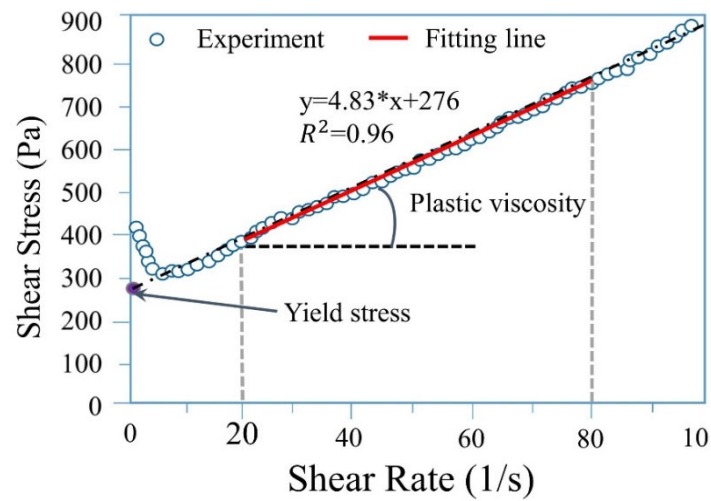

Figure 3. Test result of rheological property of the cement mortar used in this work and the fitted result by Bingham model.

## 3. Methods

### *3.1. Experimental methods*

#### *3.1.1. Direction regulation*

Due to the anisotropy of 3D printed structure, it is necessary to define the coordinate system. In this work, the moving direction of the nozzle is defined as X-axis, which means that all the printed filaments are parallel to X-axis. The direction in which the printing height increases gradually (the opposite direction of gravity) is defined as Z-axis, and the direction perpendicular to XOZ plane is defined as Y-axis, as shown in Figure 4(a). The stress state of a point in the 3D space is shown in Figure 4(b).

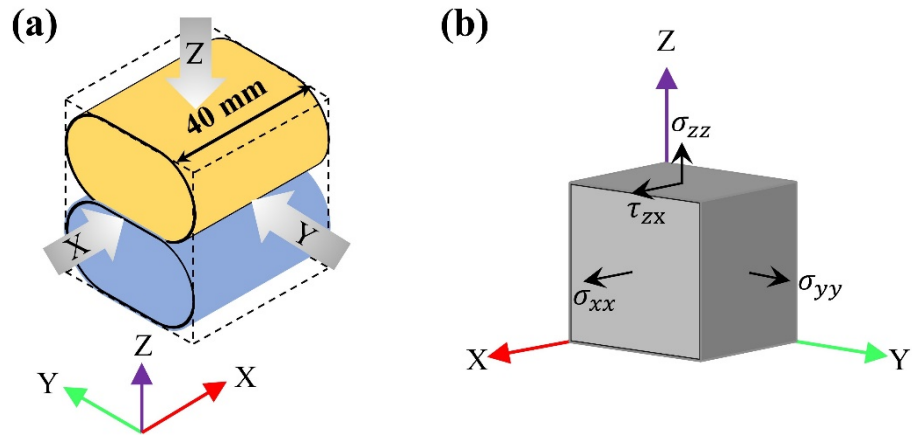

Figure 4. Orientation definition of the printed sample (a) and stress states of a point in three-dimensional space (b).

#### *3.1.2. Mechanical property test*

To study the mechanical properties of 3D printed specimen, a double-layer printed structure is prepared by continuously printing. Some specimens prepared by the traditional casting method are also used as the control-group. All specimens are cut into small pieces with different lengths (40 mm or 160 mm) to meet the requirements of various mechanical property tests. Then, the small pieces are cut into regular cubes (40 mm × 40 mm × 40 mm) and prisms (40 mm × 40 mm × 160 mm) to reduce the influence of the geometry on the mechanical properties, as shown in Figure 5.

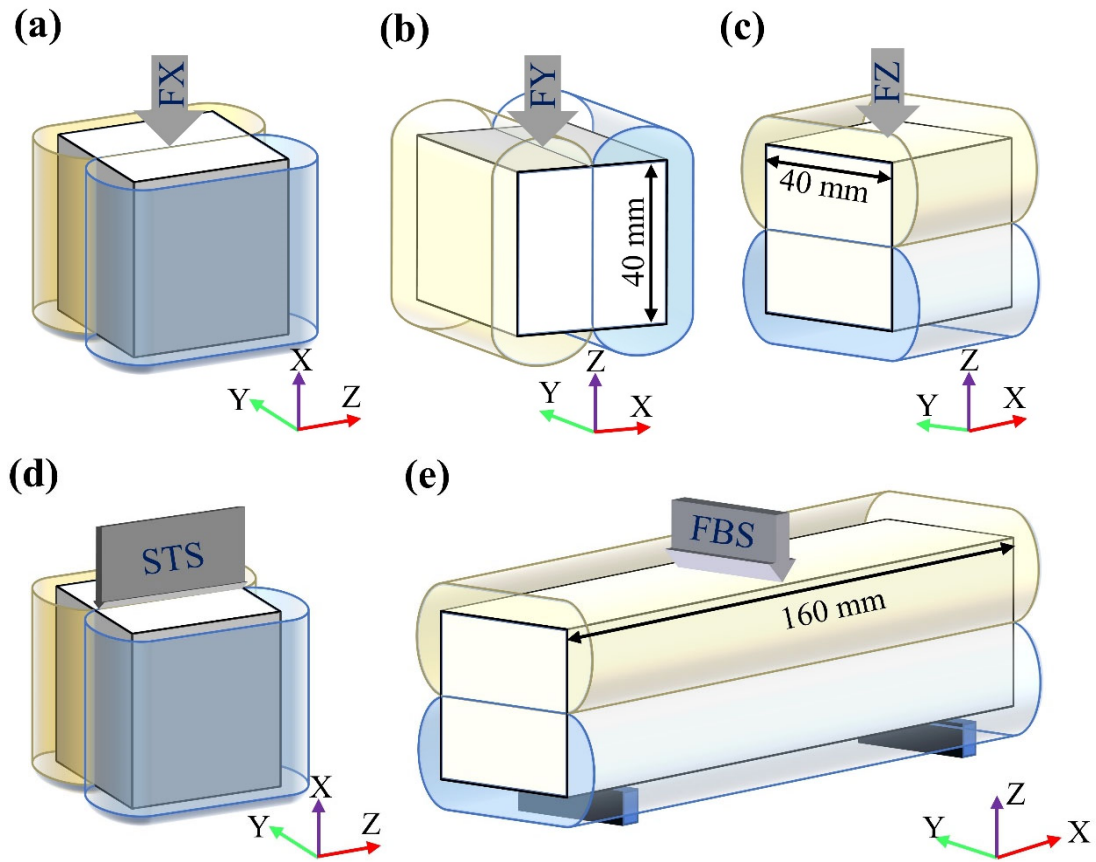

Figure 5. Double-layer printed structure used for uniaxial compressive test in X direction (a), Y direction (b) and Z direction (c) respectively, for splitting tensile test in X direction (d) and for flexural test in Z direction (e).

(1) Uniaxial compressive strength test

The small pieces with a length of 40 mm are used for uniaxial compressive strength test, as shown in Figures 5(a, b, c). Generally, the printed specimens may be loaded in X, Y and Z directions, respectively, due to the anisotropy of 3D printed specimen. For each direction, the casting specimens and printed specimens are tested in accordance with ASTM C109/C 109M-07 [32].

(2) Splitting tensile strength test

In this paper, the splitting tensile strength test targets the interlayer strength of the printed structure. The small piece with a length of 40 mm is prepared for splitting tensile strength test, as shown in Figure 5(d). The splitting test of 3D printed specimen is very similar to the cylinder splitting test. The load is applied between the strips placed in the centre on both sides of the cube specimen, and the strip position is aligned with the interlayer of the double-layer printed structure, so that the tensile failure occurs at the bonding surface of the adjacent extruded layers. All the specimens for their splitting tensile strength are tested in accordance with ASTM C 496/C 496M-04 [33]. The splitting tensile strength is calculated as follows:

$$f_{ts} = \frac{2F_{t,max}}{A\pi} = 0.637\frac{F_{t,max}}{A} \qquad (1)$$

where $f_{ts}$ is the splitting tensile strength of 3D printed sample (MPa), $F_{t,max}$ the peak load (N) in the splitting test and $A$ the bonding surface between the adjacent extruded layers.

(3) Flexural strength test

In this paper, the flexural strength test only provides the bulk strength (not the interface strength). The small piece with a length of 160 mm is used for flexural strength test, as shown in Figure 5(e). All the specimens for their flexural strength are tested in accordance with ASTM C348-18 [34], the flexural strength is calculated as follows:

$$f_{flx} = \frac{3F_{f,max} \cdot L}{2bh^2} \qquad (2)$$

where, $f_{flx}$ is the flexural strength (MPa), $F_{f,max}$ the peak load (N), $L$ the distance between two supporting points in the three-point bending test (100 mm in this work), $b$ and $h$ the width (mm) and height (mm) of the specimen, respectively.

*3.1.3. X-CT test*

X-ray computed tomography (X-CT) is an important tool to obtain the microstructure information of 3D printed structures [35]. In this work, X-ray CT scanner (Germany YXLON) is used to obtain the microstructure information of the 3D printed specimen (the small piece with a length of 40 mm) with a test resolution of 20 μm.

## 3.2. Computational-fluid-dynamics (CFD) numerical simulation method

### 3.2.1. Governing equations

In order to track the interface between the cementitious material and air, the volume of fluid (VOF) is adopted to study the flow behaviour of cementitious material during the extrusion and deposition processes. The governing equations of VOF formulations on multiphase flow are shown as follows:

Mass conservation equation:

$$\frac{\partial \rho}{\partial t} + \nabla \bullet (\rho \, \boldsymbol{u}) = 0 \tag{3}$$

Momentum conservation equation:

$$\frac{\partial \boldsymbol{u}}{\partial t} = \boldsymbol{f} + \frac{1}{\rho} \nabla \bullet T \tag{4}$$

Energy conservation equation:

$$\rho \frac{\partial}{\partial t}\left(e + \frac{u^2}{2}\right) = \rho \, \boldsymbol{f} \, \boldsymbol{u} + \nabla \bullet (\boldsymbol{u}T) - \nabla \bullet \boldsymbol{q} \tag{5}$$

where $\boldsymbol{u}$ is the velocity gradient tensor of fluid, $\boldsymbol{f}$ the volume force of unit mass, $\boldsymbol{q}$ the heat transfer coefficient, $T$ the resultant force of the surface force, $e$ the internal energy, and $\rho$ the density of fluid.

For cement-based materials, the density of material is constant in the flow process and the effect of temperature on it is not considered ($\rho$ = constant value). In addition, a shear force may be exerted on the surface of extrusion by the shaping plate during the printing process, which can be modified by defining the surface roughness of shaping plate. Surface roughness refers to the micro geometric features of the machined surface of parts, which are composed of small spacing and peak valley. It is incorporated into the shear force calculations for laminar flow behaviours. Based on the laminar flow model, the shear stress exerted by the shaping plate on the extrudate can be calculated as follows:

$$\tau_1 = \frac{\rho(v+ku)u}{\delta y} \tag{6}$$

where, $\tau_1$ is the shear stress exerted by the shaping plate on the extrudate, $v$ the kinematic viscosity, $k$ the surface roughness of shaping plate, $u$ the relative velocity, and $\delta y$ is the length scale of interest normal to the surface. Table 2 summarizes all the simulation parameters used in CFD model.

Table 2. Values of the simulation parameters.

| Parameters | Symbols | Numerical values |
|---|---|---|
| Density | $\rho$ | 2100 kg/m³ |
| Yield stress | $\sigma$ | 276 Pa |
| Plastic viscosity | $\eta_p$ | 4.83 Pa·s |
| Elastic shear modulus (EVP model) | $G$ | 200 kPa |
| Total flow rate | $U$ | 0.1 kg/s |
| Surface roughness of the shaping plate | $k$ | $1\times10^{-7}$ m |

*3.2.2. Geometry and meshing*

In this section, the filling performance of extrudate in the shaping plate apparatus and the flow behaviour of extrudate during extrusion and deposition process were investigated via computational-fluid-dynamics (CFD) numerical simulation. In the design of the shaping plate apparatus, the apparatus with different geometric parameters (such as the height of shaping plate apparatus and the offset distance of it below the nozzle) are modelled, as show in Figure 6(a). The extrusion nozzle used in CFD model is a cylindrical tube with an inner diameter D = 40 mm, a height H = 20 mm and a wall thickness of 2 mm, which is equal to that used in experiment. A "mass source" with a constant mass flow rate of 0.16 kg/s is used to continuously supple cement mortar to the nozzle and ensure the continuous printing process. The mass flow rate $U$ of "mass source" is equal to that of the printing head. The extrudate continuously fills into the shaping plate apparatus at a content speed. In addition, the shaping plate apparatus is not fixed, but moves forward along the X-axis at a constant speed *v*.

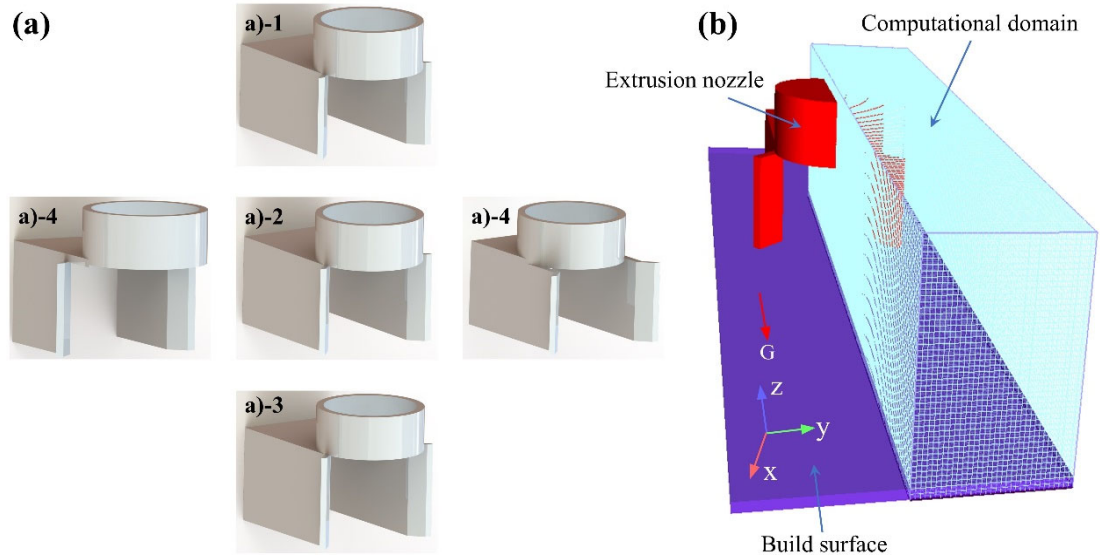

Figure 6. Shaping plate apparatus with different offset distance (in X-direction) and height (a); Simulate the filling performance of extrudate in the shaping plate apparatus (b).

A cuboid computational domain of 250 mm × 70 mm × 85 mm is used to contain all CFD models, in which the filling process is carried out. However, due to the symmetric experimental layout, a symmetric boundary condition is applied on the symmetry plane and thus only half of the domain needs to be calculated to save the calculation time, as shown in Figure 6(b). At the same time, the top, back and right sides of the computational domain are provided with pressure outlets, so that the cement mortars can flow out freely. All other boundaries of the computational domain are set as continuous boundary conditions. The whole computational domain is discretized into numerous cubic grids with 1 mm side length.

In the simulation of extrusion and deposition processes, the extrusion nozzle with and without the shaping plate apparatus are modelled, as shown in Figure 7. The shaping plate apparatus used in this stage is the apparatus that obtains best filling performance in the design stage. A cuboid computational domain of 400 mm × 70 mm

× 120 mm is used to contain all CFD models. The boundary conditions and meshing are same as that in the design stage.

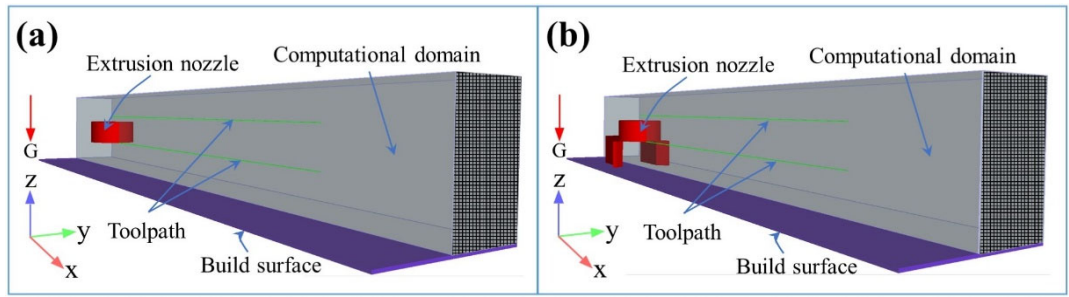

Figure 7. Geometry and computational grid of CFD model without the shaping plate (a) and with the shaping plate (b).

## 4. Design of the shaping plate apparatus and printing strategy

In the actual printer, the original nozzle is circular, as shown in Figure 8(a). Circular nozzles can adapt to the printing process under different rotation angles. However, the circular nozzle may lead to a heterogeneous migration of extrudate and thus the formation of the curved side of layers and staircase effect, which may affect the mechanical properties of 3D printed structure [36, 37]. In order to prevent the formation of staircase effect, a specially designed shaping plate apparatus is added below the nozzle, as shown as Figure 8.

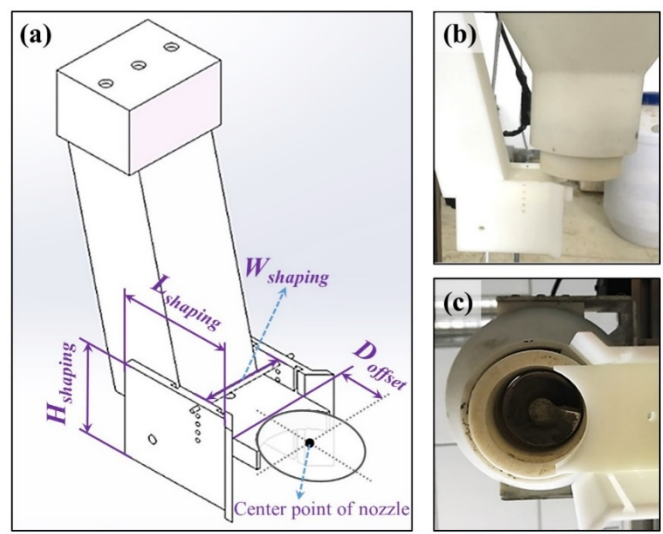

Figure 8. Design drawing of shaping plate apparatus (a) and the actual installation results of shaping plate apparatus (b, c).

## 4.1. Design of the shaping plate apparatus

As shown in Figure 9, the shaping plate apparatus is divided into six parts: (1) sleeve, (2) trapezoid ridge, (3) upper-baffle, (4) side-baffle, (5) circular through hole and (6) front-baffle. The sleeve aims at connecting the shaping plate apparatus and nozzle. The upper-baffle is used as the foundation for fixing the side-baffle through the riveting of the trapezoid ridge and the groove. The width of upper-baffle determines the width of the shaping plate apparatus, and the position of the trapezoid ridge determines the offset (in X-direction) of the shaping plate apparatus below the nozzle. The side-baffle may form a gate-shaped constraint with the upper-baffle to limit the lateral deformation of extrudate, which is the most important part of the shaping plate apparatus. The front-baffle has an angle of 135 degrees with the side-baffle, which is conducive to the smooth flow of extrudate into the shaping plate apparatus. There are many circular holes on the side-baffle, which is used to fix the side-baffle, when it is adjusted to a predetermined height.

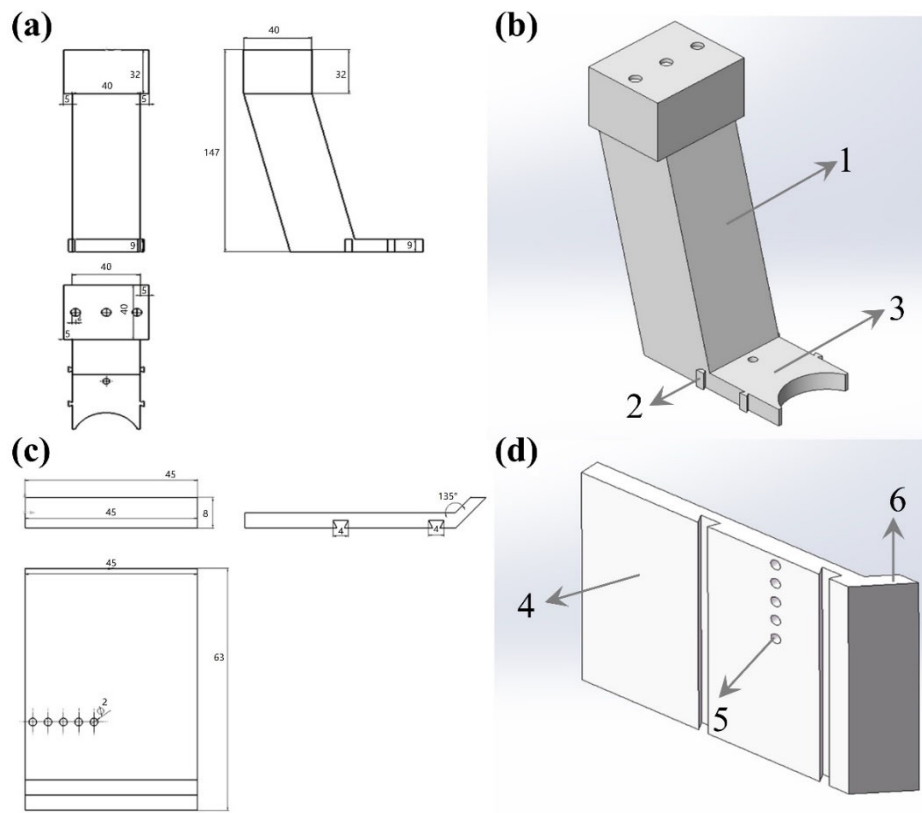

Figure 9. Design sketch of the shaping plate apparatus.

### 4.2. Optimization of the geometric parameters of the shaping plate apparatus

In the design of the shaping plate apparatus, there are several important parameters that need special attention, such as the length $L_{shaping}$, height $H_{shaping}$, width $W_{shaping}$ of the shaping plate apparatus and the offset distance (in X-direction) $D_{offset}$ of the shaping plate apparatus below the nozzle. In general, the value of $L_{shaping}$ should not be too long, especially in the structure that need to turn. The length of the apparatus may limit the minimum turning radius of structure [38]. At the same time, the value of $L_{shaping}$ should not be too short to ensure the forming of sufficient restraint space. In this work, the value of $L_{shaping}$ is set to 1.5 times the diameter of the nozzle (i.e., $L_{shaping}$ =60 mm).

Furthermore, the influences of the height (30 mm, 35 mm and 40 mm), width (35 mm, 40 mm and 45 mm) and offset distance (+R, 0 and -R) on the filling behavior of extrudate in the shaping plate apparatus were investigated via computational-fluid-dynamics (CFD) numerical simulation. In the design and optimization process of the shaping plate apparatus, the following situations should be avoided as far as possible:

(1) Printing ink overflow in the front of the shaping plate apparatus, as shown in Figure 10(a), 11 (a) and 12(c). Both low height (30 mm) and low width (35 mm) result in blockage and overflow, because the small mould space of the shaping plate apparatus is not enough to accommodate the printing ink extruded by the nozzle. Furthermore, offset distance of -R (offset R behind the nozzle along X-axis) also causes the overflow behaviour. Large offset distance of the shaping plate apparatus behind the nozzle provides an opportunity for the deformation of the printed filament. If the width of extrudate exceeds the width of the shaping plate apparatus before the shaping plate apparatus arrives, it is difficult for shaping plate apparatus to fully install the extrudate.

(2) Insufficient filling occurs in the shaping plate apparatus immediately behind the nozzle, as shown in Figure 10(c) and 11(c). Both high height and large width result in insufficient filling, because the cross-sectional area of the shaping plate apparatus is larger than that of the extrudate.

(3) The shaping plate apparatus is not fully utilized (underutilization), as shown in Figure 10(c) and 12(a). For the shaping plate apparatus with offset distance of +R, an extra room that is not filled with extrudate appears in the front of shaping plate apparatus.

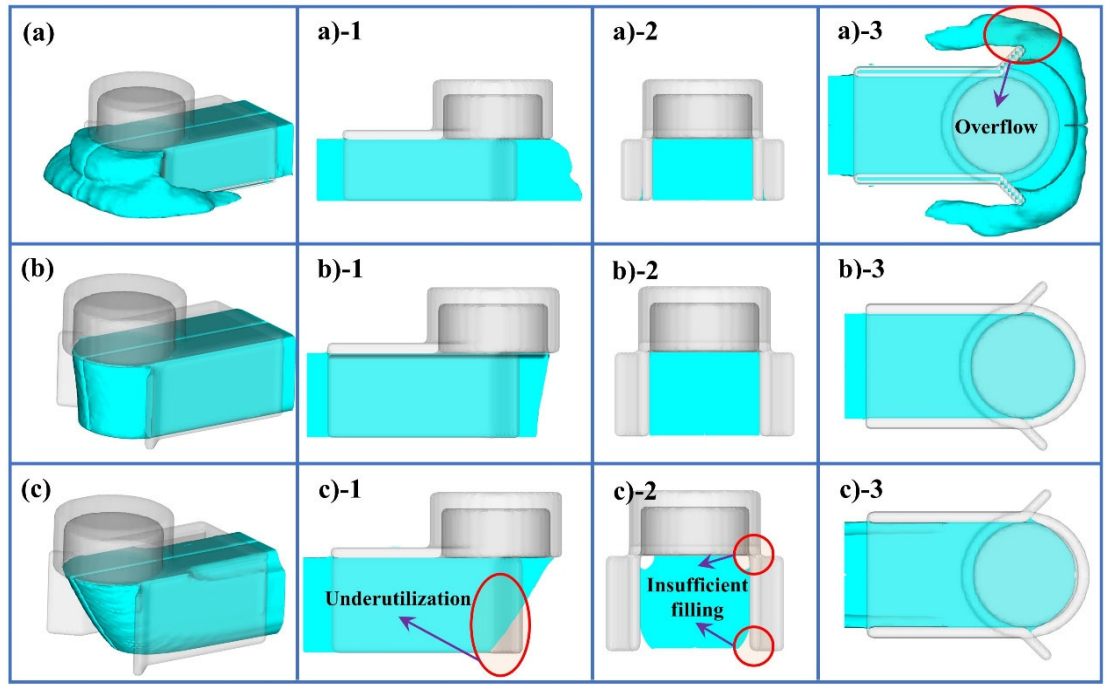

Figure 10. Filling results of extrudate in the shaping plate apparatus with a height of 30 mm (a), 35 mm (b) and 40 mm (c), respectively.

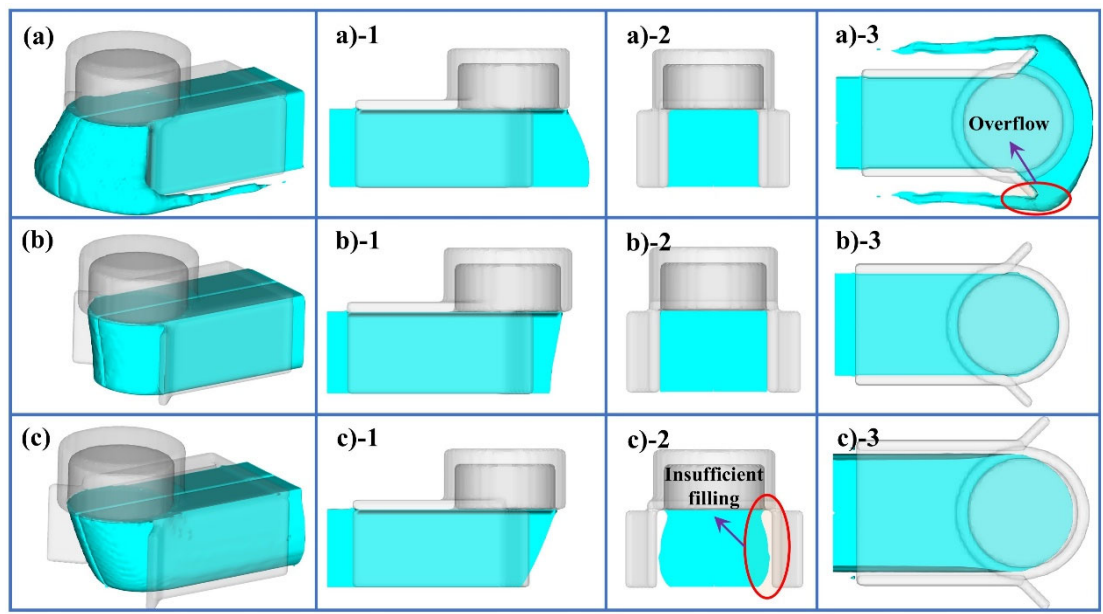

Figure 11. Filling results of extrudate in the shaping plate apparatus with a width of 35 mm (a), 40 mm (b) and 45 mm (c), respectively.

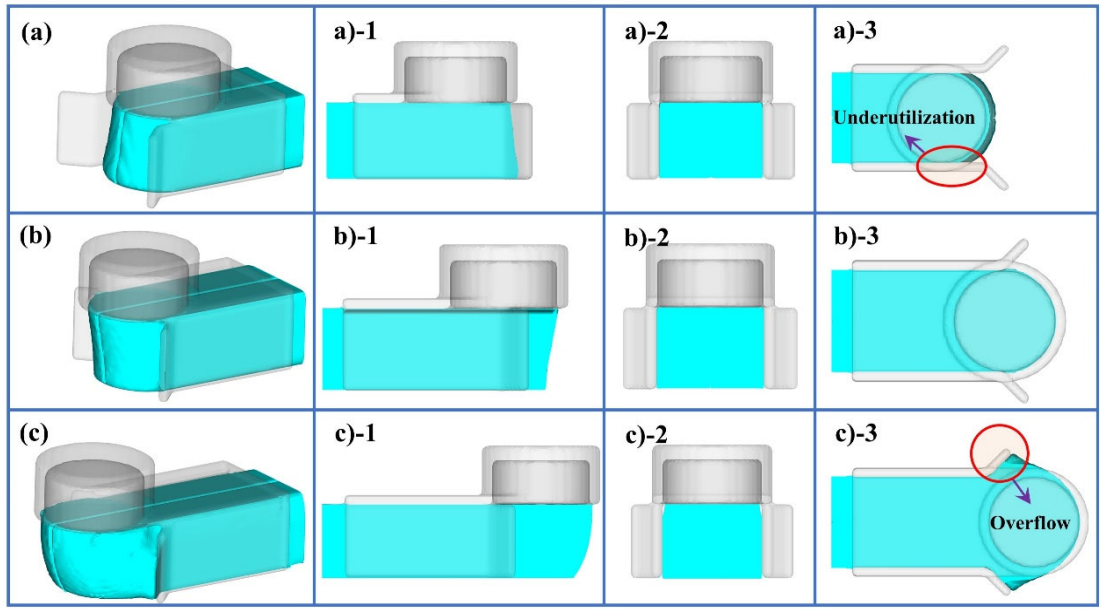

Figure 12. Filling results of extrudate in the shaping plate apparatus with a offset distance (in X-direction) of +R (a), 0 (b) and -R (c), respectively.

The filling results of extrudate in the shaping plate apparatus with different values of height, width and offset distance are summarized in Table 3. Comprehensively considering the difficulty of extrudate entering the shaping plate apparatus, the filling effect of extrudate in the shaping plate apparatus immediately behind the nozzle and the utilization of the shaping plate apparatus, the shaping plate apparatus with a height of 35 mm, a width of 40 mm and an offset distance (in X-direction) of 0 is selected in this study.

Table 3. Filling results of extrudate in the shaping plate apparatus with different values of height, width and offset distance.

| $H_{shaping}$ | $W_{shaping}$ | $D_{offset}$ | Underutilization | Insufficient filling | Overflow |
|---|---|---|---|---|---|
| 30 mm | 40 mm | 0 | | | √ |
| 35 mm | 40 mm | 0 | | | |
| 40 mm | 40 mm | 0 | √ | √ | |
| 35 mm | 45 mm | 0 | √ | √ | |
| 35 mm | 35 mm | 0 | | | √ |
| 35 mm | 40 mm | +R | √ | | |
| 35 mm | 40 mm | -R | | | √ |

## 4.3. 3D printing strategy

In 3D printing process, many factors have to be considered, such as the mass flow rate of extrudate *U*, the height *h* between the nozzle and bottom plate, the moving speed of nozzle *v*, etc. A constant mass flow rate of extrudate is critical for maintaining the continuous operation of 3D printer, which may be affected by the speed of the screw rod in the print head or the pumping pressure of the pump [15]. A constant mass flow rate of 0.16 kg/s is used in this work. In addition, the nozzle off-set between subsequent vertical layers can play a critical role. Generally, the nozzle off-set between adjacent printing layers is set as a fixed value to ensure that each printing layer has a similar geometric structure. In this work, nozzle off-set between adjacent printing layers is set as $h_{shaping}$ mm, to fully play the restraint role of the shaping plate apparatus and not to affect the movement of nozzle. Specially, for the first printed filament, the nozzle off-set between the nozzle and the bottom plate is set as $h_{shaping}$+1 mm to avoid the contacting between the shaping plate apparatus and bottom plate. The nozzle without the shaping plate apparatus also maintains the same height to form a comparison. The nozzle off-set between the current printing layer and the bottom plate may be a function of the number of printed layers and the deformation of the lower printed layers. This function may suffer an influence from different factors, such as the rheological properties, green strength, build-up rate of structure, printing time interval, etc. In this work, only double-layer structure is prepared, the deformation of the lower printed layer may be ignored. Thus, the nozzle off-set between the current printing layer and the bottom plate is calculated as follows.

$$h_{nozzle} = 1 + h_{shaping} \times n \tag{7}$$

where, $h_{nozzle}$ is the nozzle off-set between the current printing layer and the bottom plate, *n* the number of printing layers (n=1 in this study), $h_{shaping}$ the height of shaping plate apparatus. The height *h* between the nozzle and the bottom plate is important, which may influence the buildability and interlayer bonding strength of the printed structures.

In addition, the moving speed of nozzle *v* and the mass flow rate also play an important role in 3D printing [37]. For extrusion-based 3D printing, if the moving speed of nozzle *v* is not properly matched with the velocity of extrudate out of nozzle, inconsistency in the printed structures (sometimes more and sometimes less material) may occur, which may significantly influence the surface quality and mechanical properties of the printed structures [39]. As described by Rahul et al. [40], the moving speed of the nozzle *v* is related to the velocity of extrudate out of nozzle, $v_{extrudate}$, the cross-sectional area of the nozzle, *A*, and the cross-sectional area of an extruded layer $A_{layer}$. They should satisfy the following equation.

$$v = \frac{v_{extrudate} \cdot A}{A_{layer}} \qquad (8)$$

The velocity of extrudate out of nozzle, $v_{extrudate}$, is given by

$$v_{extrudate} = \frac{U}{\rho A} \qquad (9)$$

Substituting from Equation (9) in Equation (8), get

$$v = \frac{U}{\rho A_{layer}} \qquad (10)$$

where *v* is the moving speed of nozzle, *U* the mass flow rate of "mass source" or printing head, *ρ* the density of extrudate, $A_{layer}$ the cross-sectional area of an extruded layer. If the extruded element is assumed to have shape stability, then the cross-sectional area of an extruded layer from the nozzle without the shaping plate apparatus will be the same as the cross-sectional area of nozzle *A*, while the cross-sectional area of an extruded layer from the nozzle with the shaping plate apparatus will be the same as the cross-sectional area of the shaping plate apparatus $A_{shaping}$. From the stage of shaping plate apparatus design, an interesting phenomenon is found that if the cross-sectional area of the shaping plate apparatus is slightly larger than that of the nozzle, it is easy for the extrudate to flow into the shaping plate apparatus smoothly. Thus, the moving speed of nozzle with the shaping plate apparatus $v_{shaping}$ is slightly lower than that of the nozzle without the shaping plate apparatus, *v*. This may be an important conclusion for the development and application of 3D printing technology with the shaping plate apparatus (SP-3DPC). Table 4 summarizes the main parameters used in the printing process.

Table 4. Values of the printing parameters.

| Parameters | Symbols | Numerical values |
|---|---|---|
| Total flow rate | $U$ | 0.1 kg/s |
| Height between the nozzle and the bottom plate | $h$ | 36 mm |
| Outlet area of nozzle | $A$ | 1256.64 mm$^2$ |
| Outlet area of shaping plate apparatus | $A_{shaping}$ | 1400 mm$^2$ |
| Moving speed of nozzle without shaping plate apparatus | $v$ | 37.9 mm/s |
| Moving speed of nozzle with shaping plate | $v_{shaping}$ | 34 mm/s |

### *4.4. Experimental validation of the numerical model*

The virtual printing simulations provide numerical values of the pressure and constitutive stress within the printed structure, as well as a visual representation of the surface of the printed layers. The numerical results of NSP-3DPC specimen and SP-3DPC specimen are shown in Figure 13(a, c). The experimental results of NSP-3DPC specimen and SP-3DPC specimen are shown in Figure 13(b, d). It can be found that the numerical results are visually consistent with the experimental results.

Furthermore, the cross-sectional shape of the printed structures can serve as a sound basis to validate the results from the numerical simulation. As shown in Figure 13(e, f), the measured cross-sectional shape of the experimental NSP-3DPC and SP-3DPC specimens are compared with the results of the numerical simulation. It can be found that the cross-sectional shape of numerical results is in a good agreement with those of experimental results. This validates the CFD modelling approach.

## 5. Results and discussion

### 5.1. Experimental results

#### 5.1.1. Geometry of the printed filament

The cross-sectional shape of the printed structures without the shaping plate apparatus (NSP-3DPC) and with the shaping plate apparatus (SP-3DPC) are shown in Figures 13 (e) and Figures 13 (f) respectively. There is an obvious curved side of layers in NSP-3DPC specimen, which may significantly decrease the mechanical properties due to the stress concentration in the concave critical point. As shown in Figure 14(a), the top of NSP-3DPC specimen is pressed downward by the nozzle and the bottom is supported upward by the substrate layer or the bottom plate. However, there is no force at the left and right ends. Thus, the extrudate can expand freely in the direction of Y-axis, which leads to serious deformation and the formation of the curved side of layers, as shown in Figure 13(b, e).

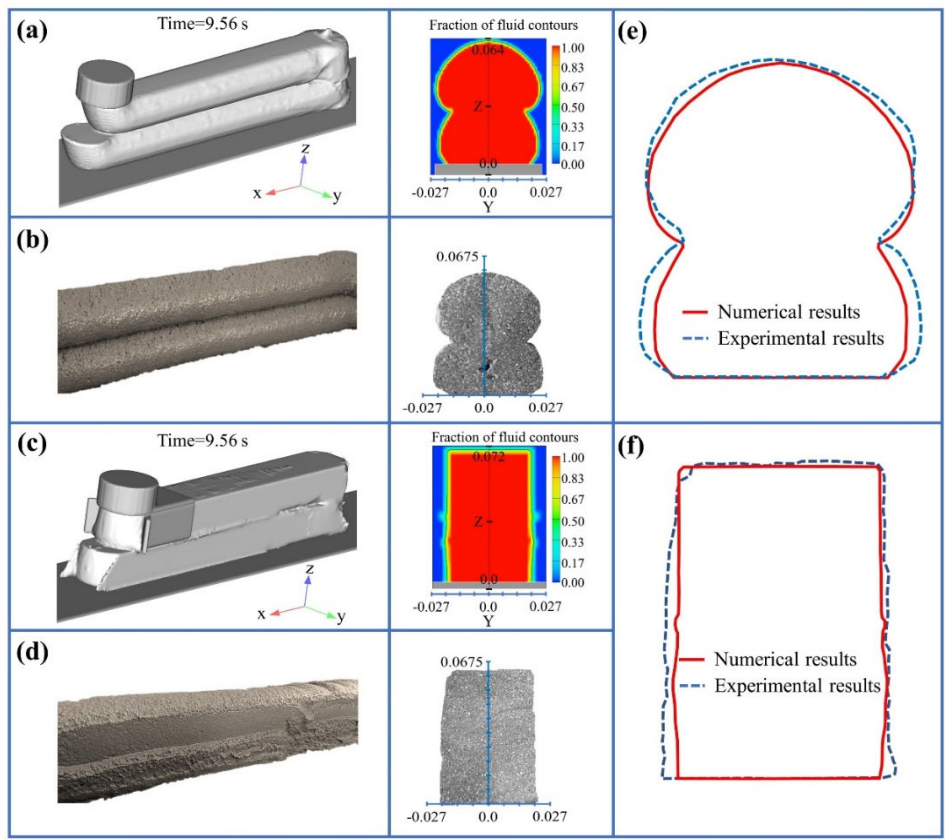

Figure 13. Numerical and experimental results of the printed structures.

Compared with the NSP-3DPC specimen, the surface finish quality of the SP-3DPC specimen is greatly improved, and the concave gap between the adjacent extruded layers disappears, as shown in Figure 13(d, f). Besides the downward pressure of the nozzle and the upward supporting force of the bottom plate, the specimen of SP-3DPC is subjected to the inward extrusion force exerted by the shaping plate, as shown in Figure 14(b). In the SP-3DPC specimen, the extrudate can only flow freely in the constraint space and continuously fill it due to the existence of hinders in four directions, so that the shape of the printed structure keeps consistent with the constraint space. Furthermore, the shaping plate also has a good smoothing effect, which eliminates the obvious interlayer between the adjacent extruded layers [29-30].

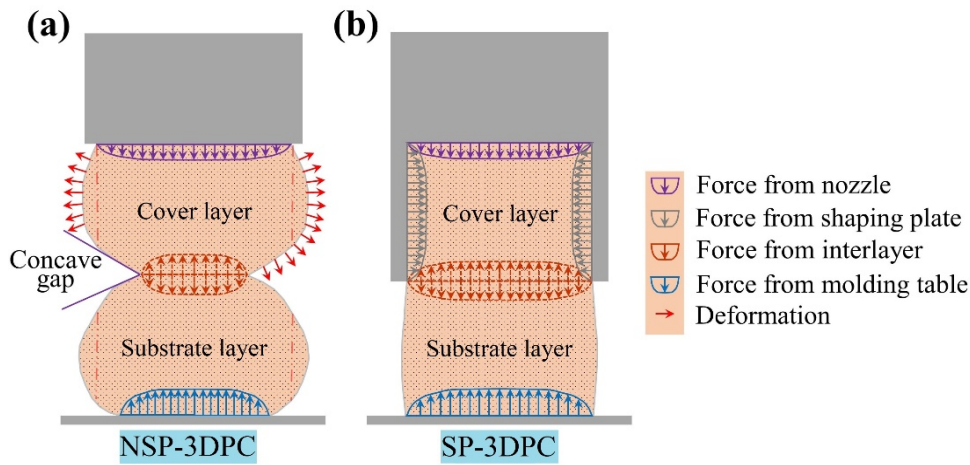

Figure 14. Schematic diagram of forces on the 3D printed sample without the shaping plate (a) and with the shaping plate (b).

*5.1.2. Mechanical property*

The test results of uniaxial compressive strength, flexural strength and splitting tensile strength are shown in Figure 15. Results indicate that the uniaxial compressive strength of NSP-3DPC specimen is lower than that of casting specimen, especially in Y direction. The uniaxial compressive strength of NSP-3DPC specimen in Y direction (23.5 MPa) is 54.1 % lower than that of the casting specimen (51.2 MPa), which is consistent with the research results of B. Panda et al [41-43]. The flexural strength of NSP-3DPC specimen (10.44 MPa) is 43.3 % lower than that of the casting specimen (18.41 MPa) and the splitting tensile strength of NSP-3DPC specimen (2.77 MPa) is 39.0 % lower than that of the casting specimen (4.54 MPa).

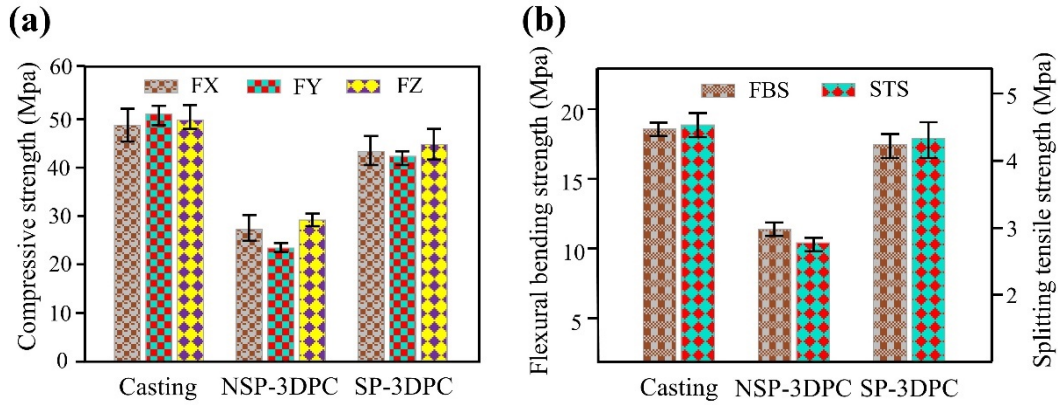

Figure 15. Uniaxial compressive strength(a), flexural bending strength (FBS) and splitting tensile strength (STS) (b) of samples.

For the NSP-3DPC specimen, the paste in the upper part of the printing layer may migrate downward under the action of gravity, which may result in more splitting/tearing cracks appearing on the upper surface of the printed structure and reduce the uniaxial compressive strength in X-axis (FX) and flexural strength. In addition, there is no constraint in the direction of Y-axis to the printed filament, thus the extrudate can flow freely along the direction of Y-axis, which may result in a large number of defects or pores remaining between the adjacent printed layers. This may be a possible explanation for the lower splitting tensile strength of NSP-3DPC specimen.

From Figure 15, it also can be seen that all of uniaxial compressive strength, flexural strength and splitting tensile strength of SP-3DPC specimen are higher than those of NSP-3DPC specimen. The compressive strength FY of SP-3DPC specimen (42.4 MPa), the flexural strength (17.23 MPa) and the splitting tensile strength (4.34 MPa) are 80.4%, 65.0%, and 56.7% higher than those of NSP-3DPC specimen, respectively. However, the mechanical properties of SP-3DPC specimen (uniaxial compressive strength, flexural strength and splitting tensile strength) are still lower than those of the casting specimen. This phenomenon is common in 3D printed cement-based materials [44, 45]. From the above analysis, it can be obviously found that using shaping plates can significantly increase the mechanical properties of 3D printed structures, but there is still a certain gap compared with the casting specimen. For SP-3DPC specimen, the extrudate may continuously fill the constraint space due to the existence of constraints in four directions, which result in a continuous densification of the printing filament. This can be used to explain why the flexural and compressive properties of SP-3DPC specimen are significantly improved. However, the compactness of SP-3DPC specimen is still far less than that of the casting specimen because the vibration in the casting process significantly increases the compactness of specimen [46]. This can be used to explain why the flexural and compressive strength of SP-3DPC specimen are lower than the casting specimen.

Furthermore, it is also found that using the shaping plate apparatus may significantly increase the splitting tensile strength of 3D printed specimen, which may be explained from the following two aspects. On the one hand, the scraping effect of the shaping plate eliminates the concave gap, increasing the fusion between the adjacent extruded layers. On the other hand, the upper baffle and side baffle (2 pieces) restricts the extrudate to expand downward along Z-axis, which increases the bonding strength between the cover layer and substrate layer. Thus, the splitting tensile strength of the SP-3DPC specimen increases significantly.

*5.1.3. Pore structure*

By X-CT slicing, the porosity images and fracture morphology images of the casting specimen, NSP-3DPC specimen and SP-3DPC specimen are shown in Figures 16 (a-c), 16 (d-f) and 16 (g-I), respectively. The statistical data of porosity in YOZ section are shown in Figure 17.

From Figure 17, it is seen that the porosity in 3D printed specimen is obviously higher than that in the casting specimen, and the average and maximum pore radius are also larger than those in the casting specimen. The casting specimen needs to be vibrated and compacted during the molding process, which eliminates the large bubbles in the specimen and increases the compactness of specimen [45]. However, 3D printed specimen lacks those vibration and compaction practices in the printing process, thus a large number of bubbles are embedded in 3D printed structures, which results in a poor compactness, as shown in Figure 16(b, c). In addition, a continuous cavity is formed in the printed filament which is caused by the spiral extrusion process, as shown in Figure 16(b). Meanwhile, it is also seen that the continuous cavity in the substrate layer is smaller than that in the cover layer, because the gravity of the cover layer compacts the substrate layer and thus compresses the continuous holes.

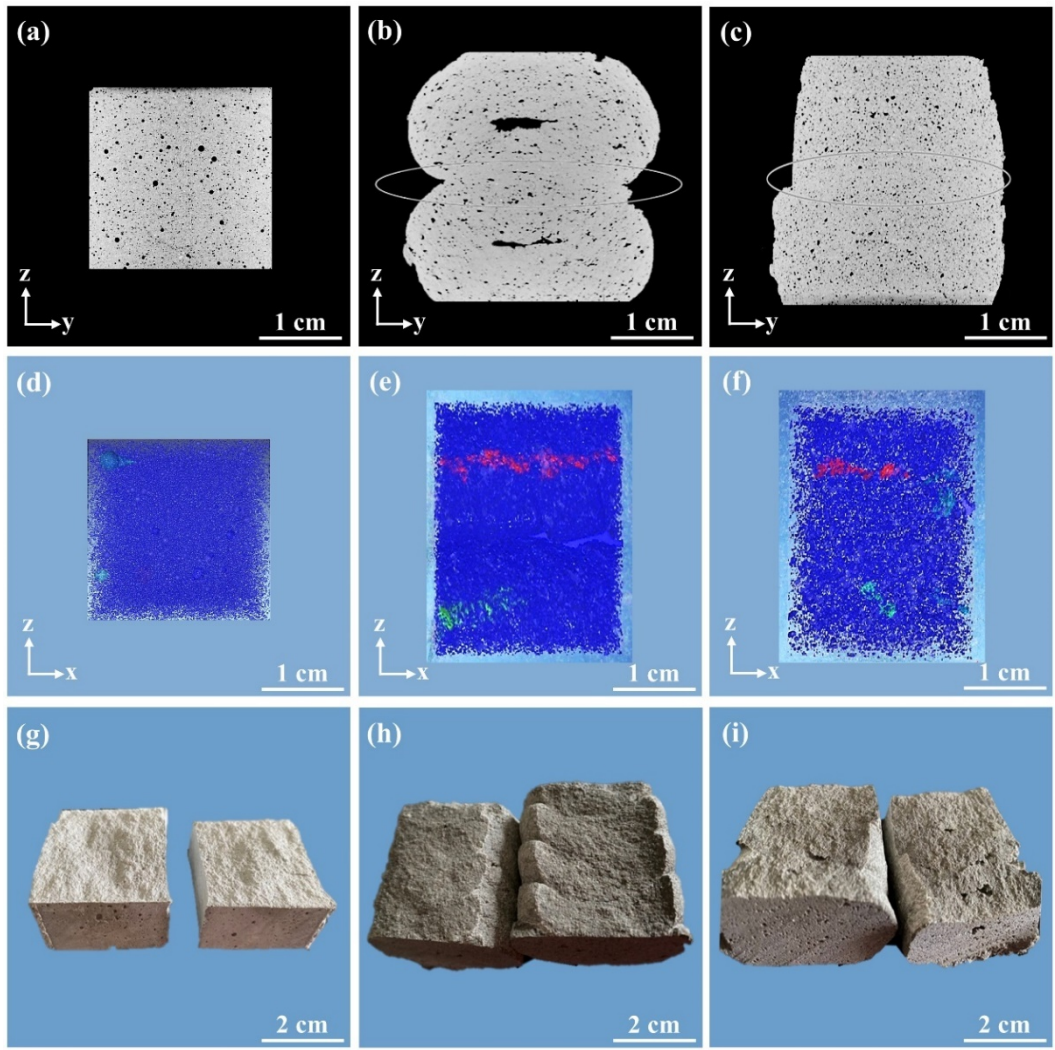

Figure 16. X-CT slice images of YOZ section of the casting sample (a), NSP-3DPC sample (b) and SP-3DPC sample (c); X-CT porosity images of XOZ section of the casting sample (d), NSP-3DPC sample (e) and SP-3DPC sample (f); Fracture photos of the casting sample (g), NSP-3DPC sample (h) and SP-3DPC sample (i).

The porosity in SP-3DPC specimen (3.79 %) is 35.5 % lower than that in the NSP-3DPC specimen (5.88 %), while the average pore radius in SP-3DPC specimen (0.29 mm) is 29.3 % lower than that in NSP-3DPC specimen (0.41 mm) too. The continuous cavity in the center of SP-3DPC specimen is not observed. The main reason for this phenomenon is that the shaping plates limit the free flow of the extrudate, and thus improve the compactness of the printed filament. This is also the reason for the significant improvements in mechanical properties of SP-3DPC specimen.

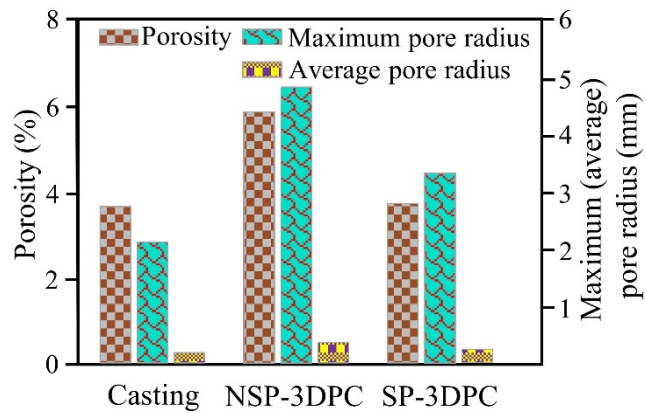

Figure 17. Porosity and pore radius of 3D-printed samples and cast sample.

The tensile failure profiles of the casting specimen, NSP-3DPC specimen and SP-3DPC specimen are shown in Figures 16(g-i), respectively. It can be seen that all of the failure surfaces are generally parallel to the loading direction. The failure surface of NSP-3DPC specimen is relatively smooth, and the fracture only converges to the interface, indicating that the weak bonding in NSP-3DPC specimen is mainly centered at the interface and the mechanical anchoring effect between the adjacent extruded layers is poor. For the casting specimen, the internal structure is relatively uniform, the crack propagation in the process of tensile failure is random, rather than concentrating in a certain plane, which results in a higher roughness of the fracture surface. For SP-3DPC specimen, the failure surface is also rough and uneven, which indicates that the weak bonding in SP-3DPC specimen is not completely centered at the interlayer.

## 5.2. Simulation results

### 5.2.1. Stress distribution in constrained region

The zz component of stress ($\sigma_{zz}$) distribution on the top surface (nozzle) and interlayer surface of 3D printed structures, the yy component of stress ($\sigma_{yy}$) distribution on the symmetry plane and the xx component of stress ($\sigma_{xx}$) distribution on the cross profile are shown in Figures 18, 19 and 20, respectively. The stress cloud can be divided into numerous strip areas which are paralleled to X-axis, Y-axis or Z-axis, and the average value of stress in each strip areas is defined as the local value of $\sigma$.

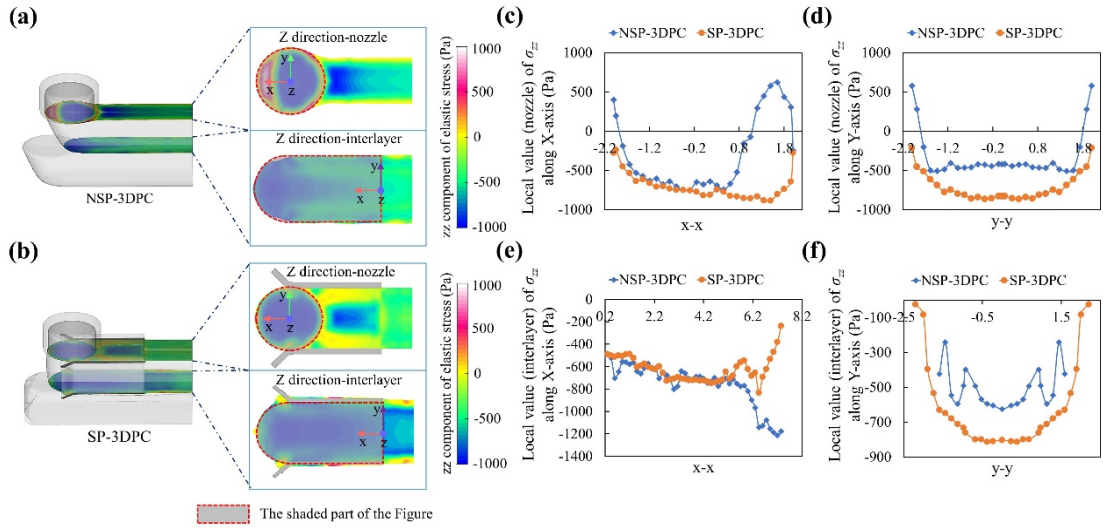

Figure 18. zz component of stress ($\sigma_{zz}$) distribution on the top surface (nozzle) and interlayer surface of PC sample with the shaping plate (a) and without the shaping plate (b); local value of $\sigma_{zz}$ on the top surface (nozzle) along X-axis (c) and Y-axis (d); local value of $\sigma_{zz}$ on the interlayer surface along X-axis (e) and Y-axis (f).

The value of $\sigma_{zz}$ on the top surface and interlayer surface are negative, as shown in Figure 18. The negative sign indicates that the direction of stress is vertical downward, meaning that compressive stress dominates on the cross section. In NSP-3DPC specimen, there is significant compressive stress concentration under the extrusion nozzle which can be transferred to the interlayer between the adjacent extruded layers. This results in obvious compressive stress concentration at the tip of the interlayer, as shown in Figures 18(a, e). Therefore, sufficient green strength is required for the printed filament to resist this stress concentration, so as to ensure the accurate printing process. If the printed filament cannot obtain sufficient green strength quickly, it may deform severely and expand along Y axis, resulting in the formation of curved side of layers. In addition, tensile stress concentration appears in the front of the top surface of NSP-3DPC specimen, as shown in Figures 18(a, c), which may be explained by that the extrudate in the front of nozzle travels longer than that at the back of nozzle during printing process.

For SP-3DPC specimen, the shaping plate apparatus form a constrained area under the extrusion nozzle, which may reshape and modify the extrudate from the nozzle. This may reduce the impact of nozzle on the printed structure. The compressive stress coming from the nozzle may not be transferred to the interlayer directly, due to the buffering effect of the constrained area. The zz component of stress ($\sigma_{zz}$) distribution on the interlayer surface of SP-3DPC specimen is more uniform than that of NSP-3DPC specimen, especially along X-axis, as shown in Figures 18(b, e). Along Y-axis, the local value of $\sigma_{zz}$ on the interlayer of SP-3DPC specimen is always lower than that of NSP-3DPC specimen, which indicates that there is more compressive stress on the interlayer of SP-3DPC specimen. This has a positive effect on the fusion between the adjacent printed layers and may be used to explain the higher splitting tensile strength of SP-3DPC specimen.

As shown in Figure 19(a, b), both values of $\sigma_{yy}$ on the symmetry plane of NSP-3DPC specimen and SP-3DPC specimen are positive. Also, the value of $\sigma_{yy}$ in the cover layer is higher than that in the substrate layer. This can be explained by that the substrate layer not only suffers from the pressure applied by the nozzle, but also bears the gravity of the cover layer. Thus, more serious volume deformation of the substrate layer occurs along Y-axis and the yy component of stress on the symmetry plane is increased. A special area where the value of $\sigma_{yy}$ is negative appears in the front of the symmetry plane of NSP-3DPC specimen, which is corresponding to the tensile stress concentration area at the front end of the top surface. This may be attributed to the inward shrinkage of the extrudate when it is stretched. Along the direction of Z-axis, the local value of $\sigma_{yy}$ on the symmetry plane of SP-3DPC specimen is always lower than that of NSP-3DPC specimen, which indicates that the possibility of expansion of SP-3DPC specimen along Y-axis is smaller than that of NSP-3DPC specimen. This phenomenon may be explained by the restraint effect of the shaping plate. The shaping plate mainly exerts an inward compressive stress on the extrudate, which hinders the expansion of extrudate in the direction of Y-axis.

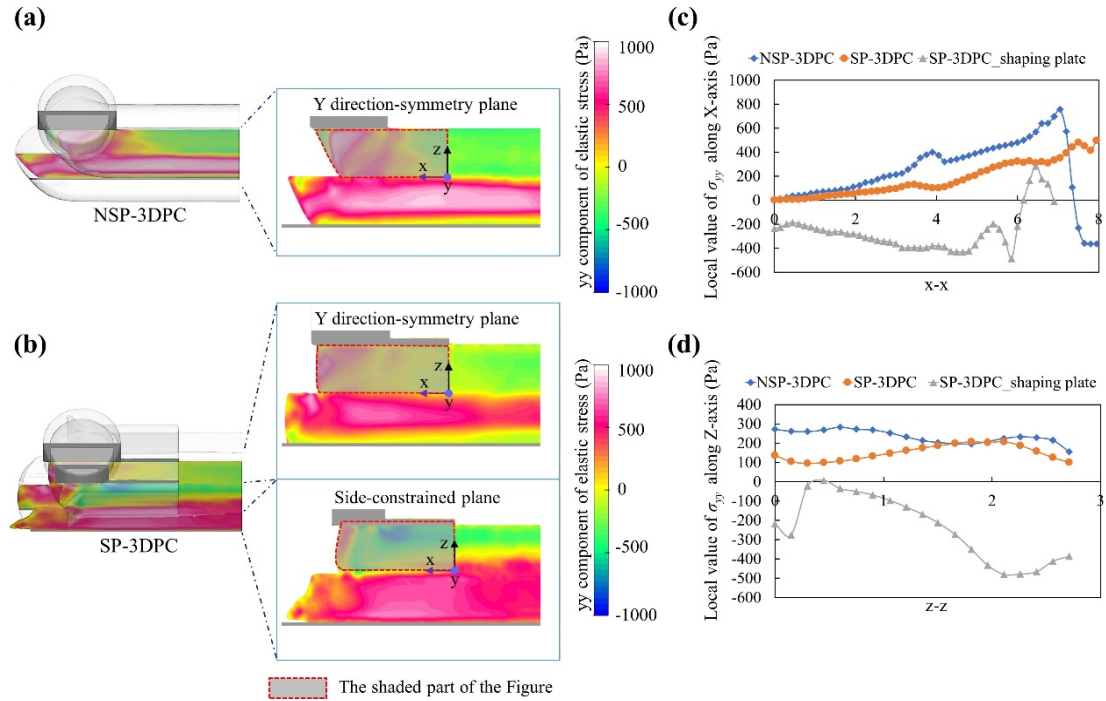

Figure 19. yy component of stress ($\sigma_{yy}$) distribution on the symmetry plane of PC sample with the shaping plate (a) and without the shaping plate (b); local value of $\sigma_{yy}$ along X-axis (c) and Z-axis (d).

The xx component of stress ($\sigma_{xx}$) distribution on the cross profile of 3D printed structures are shown in Figure 20(a, b). The average values of $\sigma_{xx}$ in both of NSP-3DPC specimen and SP-3DPC specimen are trend to zero, as shown in Figure 20(c, d). Generally, the printing structure would not be suffered from the force along the printing direction, due to the moving speed of the nozzle is always tuned right to the velocity of extrudate out of nozzle. However, the average stress trend to zero does not mean that there is no stress distribution on the cross section of the printed filament, especially in NSP-3DPC specimen. For NSP-3DPC specimen, compressive stress concentration appears in the lower part of the printed layer, while tensile stress concentration appears in the upper part of the printed layer. This may be attributed to the non-uniform migration of the extrudate. The paste in the upper part of the printing layer may migrate downward under the action of gravity, and the continuous decrease of paste in the upper part results in tensile stress. This may be a possible reason for the splitting/tearing cracks and lower flexural strength of NSP-3DPC specimen. For SP-3DPC specimen, the stress distribution in the printed structure is more uniform, because the shaping plate restricts the free expansion and migration of the extrudate.

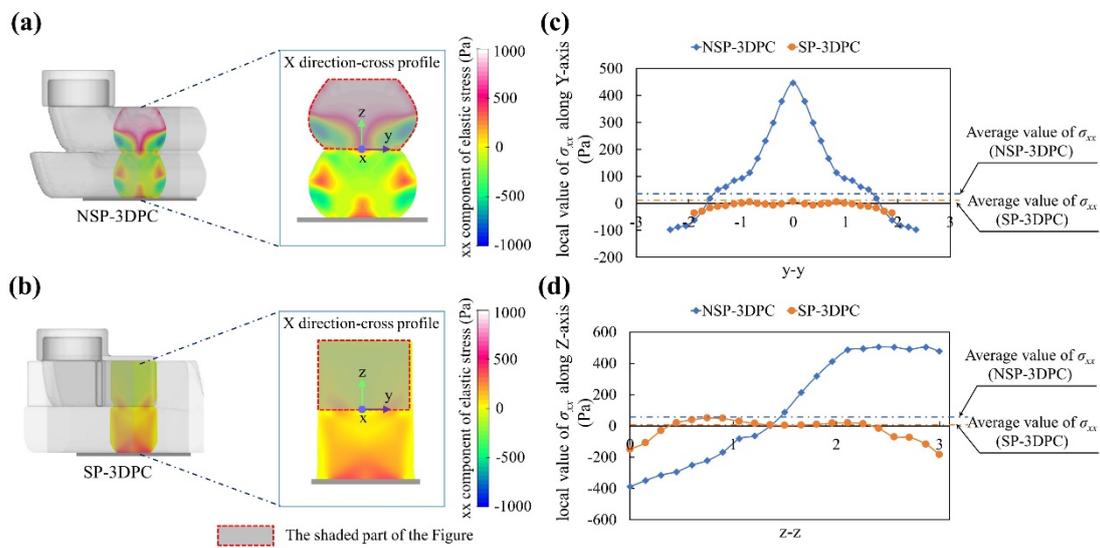

Figure 20. xx component of stress ($\sigma_{xx}$) distribution on the cross profile of PC sample with the shaping plate (a) and without the shaping plate (b); local value of $\sigma_{xx}$ along Y-axis (c) and Z-axis (d).

*5.2.2. Pressure distribution in double-layer printed structure*

The pressure $P$ distribution on the interlayer of the double-layer structure is shown in Figure 21. There is obvious pressure concentration at the front end of the interlayer in NSP-3DPC and SP-3DPC specimens. The tip of the interlayer of NSP-3DPC specimen possesses the largest pressure value, which decreases sharply at the location far from the tip and presents a single peak distribution, as shown in Figure. 21(b). However, the pressure distribution on the interface in SP-3DPC specimen is not only concentrated at the tip, but distributed in a strip shape at the front end of the interface, as shown in Figure 21(d). The pressure concentration at the front end of the interlayer is mainly attributed to the downward extrusion force. For NSP-3DPC specimen, freely expending of the extrudate consumes the extrusion force of the extrudate on the substrate layer, thus the obvious pressure concentration occurs only at the tip of interlayer where the extrusion effect is most concentrated. For SP-3DPC specimen, the shaping plate constrains the free expansion of the extrudate and decreases the loss of pressure in Y-axis, thus the moving forward of the extrusion nozzle will lead to a strip of pressure concentration at the front end of interlayer.

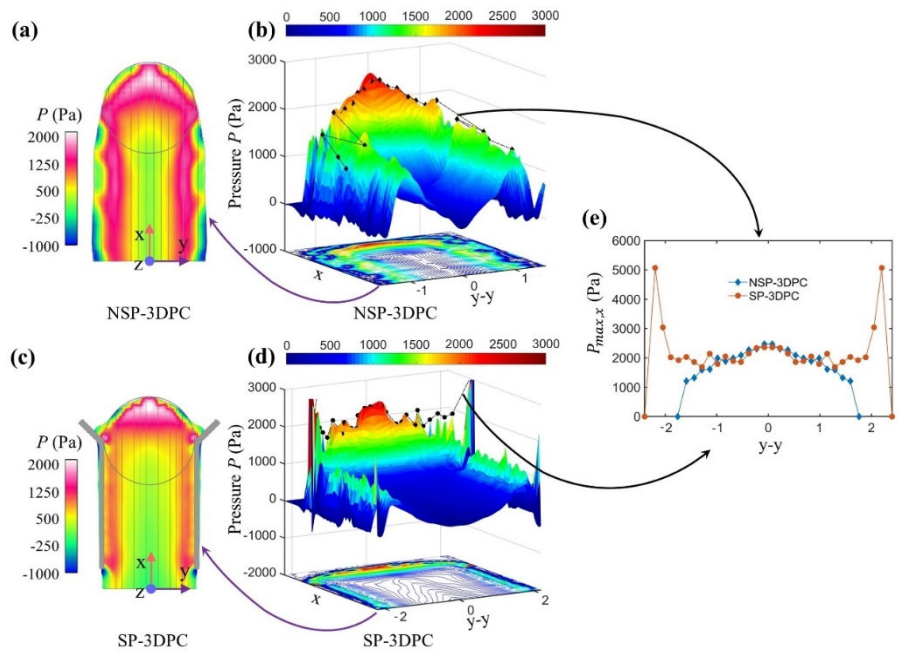

Figure 21. Contour maps and three-dimensional surface maps of the pressure $P$ distribution on the interlayer of double-layer printed specimen without the shaping plate (a and b, respectively) and with the shaping plate (c and d, respectively); Statistical result of $P_{max, x}$ (e).

The interlayer of the printed structure can be divided into numerous strip areas which are paralleled to X-axis, as shown in Figures 21(a, c), and the maximum value of pressure in each strip areas is defined as $P_{max, x}$. The curve of $P_{max, x}$ value changing with y-y value (y-y represents the distance of the statistical point from the symmetry plane of the printed structure) is shown in Figure 21(e). For NSP-3DPC specimen, the value of $P_{max,\,x}$ at the symmetry plane is the largest, and then decreases with the y-y value until the value of $P_{max,\,x}$ drops to 0 Pa at the distance of 17.5 mm from the symmetry plane. For SP-3DPC specimen, the value of $P_{max,\,x}$ near the symmetry plane fluctuates slightly, and there is a sharp peak at the distance of 21.2 mm from the symmetry plane. In general, the values of $P_{max,\,x}$ in SP-3DPC specimen are greater than that in NSP-3DPC specimen, and the distribution in SP-3DPC specimen is more uniform. The reduction of $P_{max,\,x}$ value on the interlayer can be attributed to the free expansion of the extrudate which consumes the extrusion pressure in the extrusion process. In NSP-3DPC specimen, the farther away from the symmetry plane, the stronger the free expansion of the extrudate, and the more considerably $P_{max,\,x}$ value decreases, as shown in Figure 21(e). In SP-3DPC specimen, a small amount of the extrudate will be gathered near the shaping plate, as shown in Figure 21(c, d), which exerts a large pressure on the substrate layer and results in a sharp peak of $P_{max,\,x}$ value, under the pushing force of the shaping plate.

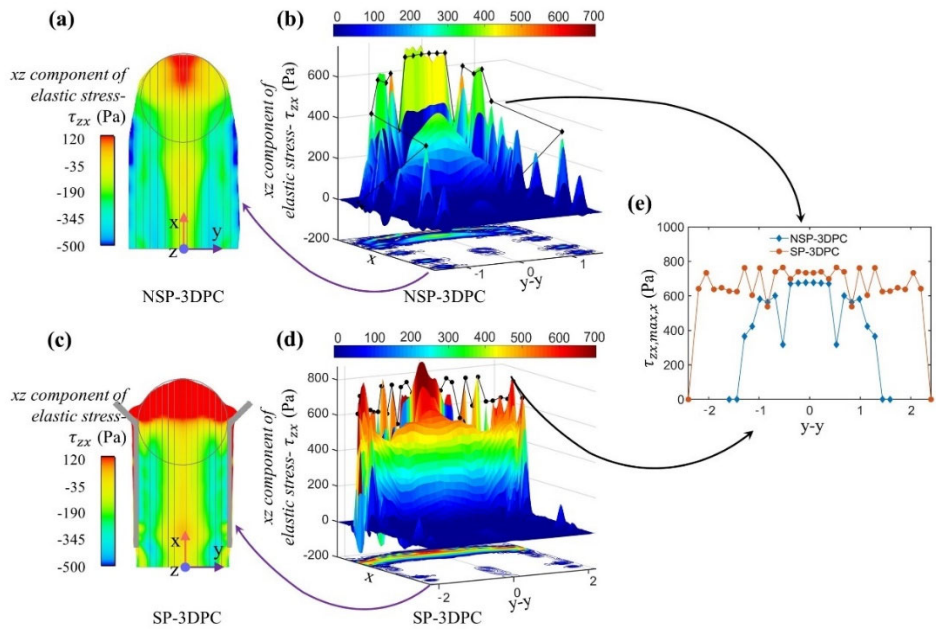

Figure 22. Contour maps and three-dimensional surface maps of the shear stress $\tau_{zx}$ distribution on the interlayer of double-layer printed specimen without the shaping plate (a and b, respectively) and with the shaping plate (c and d, respectively); Statistical result of $\tau_{zx\,max,\,x}$ (e).

The extrusion force of extrudate not only leads to the pressure concentration on the interface, but also affects the shear stress distribution on the interlayer. The distribution of shear stress $\tau_{zx}$ on the interlayer of the double-layer printed structures is shown in Figure 22. The shear stress $\tau_{zx}$ has a similar variation trend as the pressure $P$. Compared with NSP-3DPC specimen, the value of $\tau_{zx, max, x}$ in SP-3DPC specimen is larger, and its distribution range is wider.

In conclusion, compared with NSP-3DPC specimen, SP-3DPC specimen exudes higher pressure $P$ and higher shear stress $\tau_{zx}$ on the interlayer, which increases the bonding strength between the adjacent extruded layers. This can be used to explain why SP-3DPC specimen has higher flexural strength and splitting tensile strength.

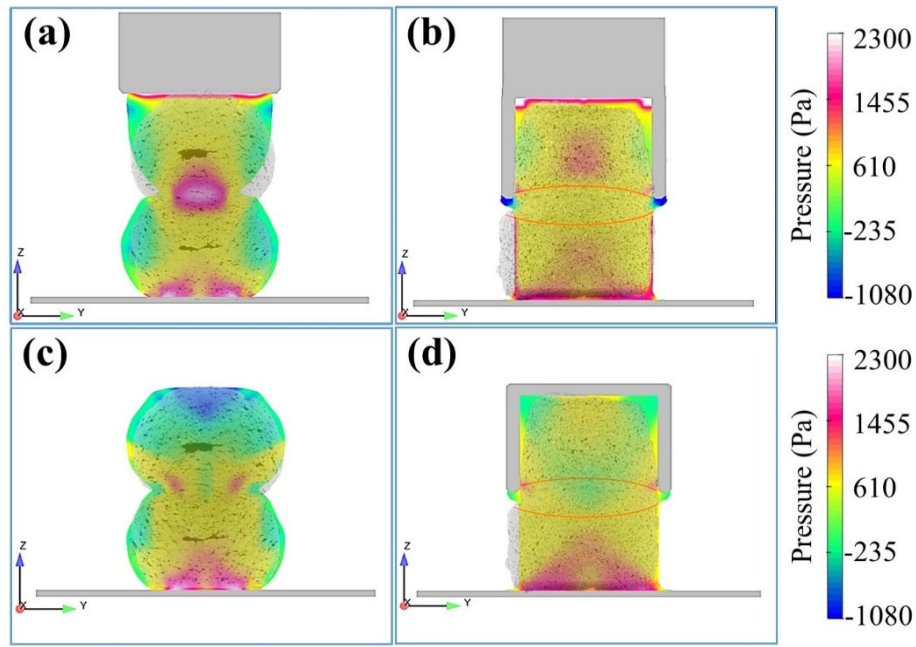

Figure 23. Pressure distribution on the cross-section A (YOZ section, which directly below the extrusion nozzle) of NSP-3DPC specimen (a) and SP-3DPC specimen (b); Pressure distribution on the cross-section B (YOZ section, which misses the extrusion nozzle but doesn't exceed the constraint region of shaping plate) of NSP-3DPC specimen (c) and SP-3DPC specimen (d).

The pressure distributions on the cross-section A (YOZ section, which directly below the extrusion nozzle) of NSP-3DPC specimen and SP-3DPC specimen, are shown in Figures 23(a, b). The pressure distributions on the cross-section B (YOZ section, which misses the extrusion nozzle but doesn't exceed the constraint region of shaping plate of NSP-3DPC specimen and SP-3DPC specimen, are shown in Figures 23(c, d). For NSP-3DPC specimen, both the pressure distributions on cross-section A and cross-section B are heterogeneous, due to the gourd-shaped structure of NSP-3DPC specimen. For SP-3DPC specimen, the configuration of specimen is more regular, because the shaping plate restricts the lateral deformation of the printed filament. Thus, both the pressure distributions on cross-section A and cross-section B are more uniform.

As described in Section 5.1.3, continuous cavities caused by the spiral extrusion process are formed in NSP-3DPC specimen, as shown in Figure 15(b). However, these continuous cavities are absent in SP-3DPC specimen. This phenomenon can be explained by the pressure distribution on the cross-section A of SP-3DPC specimen. In SP-3DPC specimen, the upper surface is blocked by the extrusion nozzle and upper baffle; the lower surface is supported by the substrate layer or bottom plate; the left and right surfaces are constrained by the shaping plates. Thus, a pressure concentration occurs in the centre of the printed filament, as shown in Figure 23(b), which increases the compactness of the printed filament and eliminates the continuous cavities.

## 6. Conclusions

In this work, the mechanical properties, microstructure and internal stress and pressure distribution of 3D printed structure with/without the shaping plate (NSP-3DPC/SP-3DPC specimens) are investigated via mechanical property tests, X-ray micro-computed tomography (CT) and single-phase computational fluid dynamics (CFD) simulation. 3D printed structure without shaping plate (NSP-3DPC specimen) is set as the control group. The following conclusions are obtained:

(1) A slightly larger outlet area of the shaping plate apparatus than that of the nozzle can easily ensure a smoother flow of the extrudate into the shaping plate apparatus, which may slightly reduce the printing speed. This may be an important conclusion for the development and application of 3D printing technology with a shaping plate apparatus.

(2) The shaping plate apparatus can significantly improve the mechanical properties of the 3D printed structure. Compared with NSP-3DPC specimen, SP-3DPC specimen has higher uniaxial compressive strength, flexural strength and splitting tensile strength. The splitting tensile strength and flexural strength of SP-3DPC specimen are 56.7% and 65.0% higher than those of NSP-3DPC specimen, respectively.

(3) The shaping plate apparatus constrains the free expansion of the extrudate, restrains the deformation of the printed filament, and ensures the geometric accuracy of the printed filament. Furthermore, the shaping plates act as a shovel to mechanically comb excess material along the side of the nozzle, significantly improving the surface finish quality of the printed structure.

(4) In the printing process, the shaping plate exerts an inward pressure on the printed filament, which changes the stress distribution inside it. Compared with NSP-3DPC specimen, SP-3DPC specimen has larger and more evenly distributed internal stress. This ensures the compactness and uniformity of the internal structure, reduces the porosity and average pore size of specimen, and significantly improves the uniaxial compressive strength.

(5) Compared with NSP-3DPC specimen, SP-3DPC specimen exudes higher pressure $P$ and shear stress $\tau_{zx}$ on the interlayer, which increases the bonding strength between the adjacent extruded layers and significantly increases the splitting tensile strength of the printed structure.

(6) An obvious stress concentration occurs at the center of the printed filament in SP-3DPC specimen. This eliminates the continuous holes, reduces the porosity and the maximum pore radius, and significantly improves the mechanical properties of the printed structure.

**CRediT authorship contribution statement**

Tinghong Pan: Conceptualization, Methodology, Investigation, Formal analysis, Writing - Original draft, Review and Editing. Huaijin Teng: Conceptualization, Methodology, Investigation, Formal analysis. Hengcheng Liao: Conceptualization, Methodology, Writing – Review and Editing, Supervision, Funding acquisition.

Yaqing Jiang: Conceptualization, Methodology, Writing – Review and Editing, Supervision, Funding acquisition. Chunxiang Qian: Supervision, Funding acquisition. Yu Wang: Methodology, Writing – Review and Editing.

**Declaration of competing interest**

The authors declare that they have no known competing financial interests or personal relationships that could have appeared to influence the work reported in this paper.

**Acknowledgments**

This work was supported by the National Natural Science Foundation of China [grant numbers 51738003, 11772120]; and the Primary Research & Development Plan of Jiangsu Province [grant number BE2016187].

**References**

[1] N. Labonnote, A. Rønnquist, B. Manum and P. Rüther, Additive construction: State-of-the-art, challenges and opportunities, Autom. Constr. 72(2016) 347-366.

[2] L. He, W.T. Chow and H. Li, Effects of interlayer notch and shear stress on interlayer strength of 3D printed cement paste, Addit. Manuf. 36(2020) 101390.

[3] Z. Liu, M. Li, Y.W.D. Tay, Y. Weng, T.N. Wong and M.J. Tan, Rotation nozzle and numerical simulation of mass distribution at corners in 3D cementitious material printing, Addit. Manuf. 34(2020) 101190.

[4] E. Lloret, A.R. Shahab, M. Linus, R.J. Flatt, F. Gramazio, M. Kohler and S. Langenberg, Complex concrete structures: Merging existing casting techniques with digital fabrication, Comput. Aided Des. 60(2015) 40-49.

[5] V.N. Nerella, M. Krause and V. Mechtcherine, Direct printing test for buildability of 3D-printable concrete considering economic viability, Autom. Constr. 109(2020) 102986.

[6] V. Mechtcherine, V.N. Nerella, F. Will, M. Näther, J. Otto and M. Krause, Large-scale digital concrete construction – CONPrint3D concept for on-site, monolithic

3D-printing, Autom. Constr. 107(2019) 102933.

[7] A. Anton, L. Reiter, T. Wangler, V. Frangez, R.J. Flatt and B. Dillenburger, A 3D concrete printing prefabrication platform for bespoke columns, Autom. Constr. 122(2021) 103467.

[8] C. Gosselin, R. Duballet, P. Roux, N. Gaudillière, J. Dirrenberger and P. Morel, Large-scale 3D printing of ultra-high performance concrete – a new processing route for architects and builders, Mater. Des. 100(2016) 102-109.

[9] M.T. Souza, I.M. Ferreira, E. Guzi De Moraes, L. Senff and A.P. Novaes De Oliveira, 3D printed concrete for large-scale buildings: An overview of rheology, printing parameters, chemical admixtures, reinforcements, and economic and environmental prospects, J. Build. Eng. 32(2020) 101833.

[10] H. Alhumayani, M. Gomaa, V. Soebarto and W. Jabi, Environmental assessment of large-scale 3D printing in construction: A comparative study between cob and concrete, J. Clean. Prod. 270(2020) 122463.

[11] D. Heras Murcia, M. Genedy and M.M. Reda Taha, Examining the significance of infill printing pattern on the anisotropy of 3D printed concrete, Constr. Build. Mater. 262(2020) 120559.

[12] S. Yu, H. Du and J. Sanjayan, Aggregate-bed 3D concrete printing with cement paste binder, Cem. Concr. Res. 136(2020) 106169.

[13] D. Lowke, D. Talke, I. Dressler, D. Weger, C. Gehlen, C. Ostertag and R. Rael, Particle bed 3D printing by selective cement activation – Applications, material and process technology, Cem. Concr. Res. 134(2020) 106077.

[14] D. Lowke, E. Dini, A. Perrot, D. Weger, C. Gehlen and B. Dillenburger, Particle-bed 3D printing in concrete construction – Possibilities and challenges, Cem. Concr. Res. 112(2018) 50-65.

[15] A. Perrot, D. Rangeard, 3D Printing in Concrete: Techniques for Extrusion/Casting, ISTE Ltd and John Wiley & Sons Inc., (2019) 41-72.

[16] V. Mechtcherine, F.P. Bos, A. Perrot, W.R.L. da Silva, V.N. Nerella, S. Fataei, R.J.M. Wolfs, M. Sonebi and N. Roussel, Extrusion-based additive manufacturing with cement-based materials – Production steps, processes, and their underlying

physics: A review, Cem. Concr. Res. 132(2020) 106037.

[17] R.J. Wolfs, T.A. Salet and N. Roussel, Filament geometry control in extrusion-based additive manufacturing of concrete: The good, the bad and the ugly, Cem. Concr. Res. 150(2021) 106615.

[18] N. Roussel, Rheological requirements for printable concretes, Cem. Concr. Res. 112(2018) 76-85.

[19] N. Roussel, J. Spangenberg, J. Wallevik and R. Wolfs, Numerical simulations of concrete processing: From standard formative casting to additive manufacturing, Cem. Concr. Res. 135(2020) 106075.

[20] P.F. Yuan, Q. Zhan, H. Wu, H.S. Beh and L. Zhang, Real-time toolpath planning and extrusion control (RTPEC) method for variable-width 3D concrete printing, J. Build. Eng. (2021) 103716.

[21] J. Kruger and G. van Zijl, A compendious review on lack-of-fusion in digital concrete fabrication, Addit. Manuf. 37(2021) 101654.

[22] F. Bester, M. van den Heever, J. Kruger and G. van Zijl, Reinforcing digitally fabricated concrete: a systems approach review, Addit. Manuf. 37(2021) 101737.

[23] R.A. Buswell, W.L. De Silva, S.Z. Jones and J. Dirrenberger, 3D printing using concrete extrusion: A roadmap for research, Cem. Concr. Res. 112(2018) 37-49.

[24] W. Lao, M. Li, T.N. Wong, M.J. Tan and T. Tjahjowidodo, Improving surface finish quality in extrusion-based 3D concrete printing using machine learning-based extrudate geometry control, Virtual Phys. Prototyp. 15(2020) 178-193.

[25] W. Lao, M. Li and T. Tjahjowidodo, Variable-geometry nozzle for surface quality enhancement in 3D concrete printing, Addit. Manuf. 37(2021) 101638.

[26] S.C. Paul, G.P. van Zijl, M.J. Tan and I. Gibson, A review of 3D concrete printing systems and materials properties: Current status and future research prospects, Rapid Prototyping J. (2018).

[27] W. Lao, D.Y.W. Tay, D. Quirin and M.J. Tan, The effect of nozzle shapes on the compactness and strength of structures printed by additive manufacturing of concrete, in: Proceedings of the 3rd International Conference on Progress in Additive Manufacturing (Pro-AM 2018), Singapore, 2018, pp. 14-17.

[28] K. Manikandan, X. Jiang, A.A. Singh, B. Li and H. Qin, Effects of Nozzle Geometries on 3D Printing of Clay Constructs: Quantifying Contour Deviation and Mechanical Properties, Procedia Manuf. 48(2020) 678-683.

[29] B. Khoshnevis and R. Dutton, Innovative Rapid Prototyping Process Makes Large Sized, Smooth Surfaced Complex Shapes in a Wide Variety of Materials, Mater. technol. (New York, N.Y.) 13(1998) 53-56.

[30] B. Khoshnevis, D. Hwang, K. Yao and Z. Yeh, Mega-scale fabrication by contour crafting, Int. J. Ind. Syst. Eng. 1(2006) 301-320.

[31] P. Carneau, R. Mesnil, O. Baverel and N. Roussel, Layer pressing in concrete extrusion-based 3D-Printing: experiments and analysis, submitted to Cement Concr. Res.

[32] ASTM. Standard Test Method for Compressive Strength of Hydraulic Cement Mortars (Using 2-in. or [50-mm] Cube Specimens); ASTM C 109/C109M-07; ASTM International: West Conshohocken, PA, USA, 2008.

[33] ASTM. Standard Test Method for Splitting Tensile Strength of Cylindrical Concrete Specimens; ASTM C 496/C 496M-04; ASTM International: West Conshohocken, PA, USA, 2004.

[34] ASTM. Standard Test Method for Flexural Strength of Hydraulic Cement Mortars; ASTM C348-18; ASTM International: West Conshohocken, PA, USA, 2019.

[35] J. Kruger, A. du Plessis and G. van Zijl, An investigation into the porosity of extrusion-based 3D printed concrete, Addit. Manuf. 37(2021) 101740.

[36] P. Shakor, S. Nejadi and G. Paul, A Study into the Effect of Different Nozzles Shapes and Fibre-Reinforcement in 3D Printed Mortar, Materials. 12(2019) 1708.

[37] F. Bos, R. Wolfs, Z. Ahmed and T. Salet, Additive manufacturing of concrete in construction: potentials and challenges of 3D concrete printing, Virtual Phys. Prototyp. 11(2016) 209-225.

[38] W. McGee, T.Y. Ng, K. Yu and V.C. Li, Extrusion Nozzle Shaping for Improved 3DP of Engineered Cementitious Composites (ECC/SHCC), in: Second RILEM International Conference on Concrete and Digital Fabrication, (2020) 916-925.

[39] Y.W.D. Tay, B. Panda, S.C. Paul, N.A. Noor Mohamed, M.J. Tan and K.F. Leong,

3D printing trends in building and construction industry: a review, Virtual Phys. Prototyp. 12(2017) 261-276.

[40] A.V. Rahul, M. Santhanam, H. Meena and Z. Ghani, 3D printable concrete: Mixture design and test methods, Cem. Concr. Compos. 97(2019) 13-23.

[41] B. Panda, S. Chandra Paul and M. Jen Tan, Anisotropic mechanical performance of 3D printed fiber reinforced sustainable construction material, Mater. Lett. 209(2017) 146-149.

[42] B. Panda, S.C. Paul, N.A.N. Mohamed, Y.W.D. Tay and M.J. Tan, Measurement of tensile bond strength of 3D printed geopolymer mortar, Measurement. 113(2018) 108-116.

[43] B. Panda, N.A. Noor Mohamed, S. Chana Paul, G. Bhagath Singh, M.J. Tan and B. Šavija, The Effect of Material Fresh Properties and Process Parameters on Buildability and Interlayer Adhesion of 3D Printed Concrete, Materials. 12(2019) 2149.

[44] C. Joh, J. Lee, J. Park and I. Yang, Buildability and Mechanical Properties of 3D Printed Concrete, Materials. 13(2020) 4919.

[45] R. Rao, Q. Deng, J. Fu, C. Liu, X. Ouyang and Y. Huang, Improvement of mechanical strength of recycled blend concrete with secondary vibrating approach, Constr. Build. Mater. 237(2020) 117661.

[46] J. Huang, J. Pei, Y. Li, H. Yang, R. Li, J. Zhang and Y. Wen, Investigation on aggregate particles migration characteristics of porous asphalt concrete (PAC) during vibration compaction process, Constr. Build. Mater. 243(2020) 118153.